\documentclass[11pt]{article}

\def\be{\begin{equation}}
\def\ee{\end{equation}}
\def\ba{\begin{eqnarray}}
\def\ea{\end{eqnarray}}
\def\la{\langle}
\def\ra{\rangle}
\def\a{\alpha}
\def\b{\beta}

\usepackage{graphicx}
\begin{document}
\begin{titlepage}
\vspace{4cm}
\begin{center}{\Large \bf Thermal Entanglement of bosonic modes}\\
\vspace{1cm}M. Asoudeh\footnote{email:asoudeh@mehr.sharif.edu},
\hspace{0.5cm} \\
\vspace{1cm} Department of Physics, Sharif University of Technology,\\
P.O. Box 11365-9161,\\ Tehran, Iran
\end{center}
\vskip 3cm
\begin{abstract}
We study the change of entanglement under general linear
transformation of modes in a bosonic system and determine the
conditions under which entanglement can be generated under such
transformations. As an example we consider the thermal entanglement
between the vibrational modes of two coupled oscillators and
determine the temperature above which quantum correlations are
destroyed by thermal fluctuations.
\end{abstract}
\end{titlepage}
\vskip 3cm

\section{Introduction}

Consider two ions in an ion trap, or two atoms in a solid vibrating
around their equilibrium positions. We ask how much pure quantum
correlation or entanglement exists between the vibrational modes of
these two ions or atoms at a given temperature? How much we can
raise the temperature before the quantum correlation between the
ions is destroyed? In a more general context, we can ask the degree
of entanglement of two bosonic modes in a many body system. This
property known as thermal entanglement has been intensively studied
mostly for the spin degrees of freedom of atoms or ions in the past
two or three years \cite{niel, arn, ocw}. The reason for this
restriction has been threefold. The first is that a calculable
measure of entanglement when the two systems are in a mixed state
has been known only for two dimensional systems \cite{woo}. The
second reason has been obviously the interest in two dimensional
systems as representatives of qubits in quantum computers
\cite{nielchuang}. The third is that spin systems are the prototype
of interacting many body localized fermions and it has been
desirable to see what happens to entanglement when a spin system
undergoes quantum phase transition \cite{ost}.\\ While the pursuit
of this problem for intermediate dimensions has been impossible (for
the lack of a measure of entanglement) or of little interest, all
the above three motivations exist also for the other extreme of
dimensionality, namely systems with continuous degrees of freedom.
First, for a class of states called symmetric Gaussian states, a
closed formula exist for their entanglement of
formation\cite{cirac}, second, systems of continuous degrees of
freedom are of wide interest as candidates for the implementation of
quantum computing \cite{braun} and third, in condensed matter
systems, the continuous degrees of freedom like the vibrational
modes of localized atoms in a crystal or the bosonic modes of a
system of identical particles can also be
entangled. The same phenomenon is of interest in bosonic field theories \cite{ved}. \\
When dealing with systems of identical particles, it is the modes
and not the particles which should be treated as subsystems
\cite{enk} and one can quantify the entanglement of these
subsystems in a second quantized approach\cite{enk, gitt, shu,
zan, ved}. In this case entanglement changes by
redefining the modes, a non-local operation on subsystems. \\
In this paper we study a system composed of identical spinless
bosons subject to a free quadratic hamiltonian and study the thermal
entanglement properties of different arbitrary modes of this system.
This problem will also be of relevance when we want to understand
how thermal fluctuations will affect the efficiency
of protocols based on gaussian states \cite{cerf}.  \\
So we consider a hamiltonian of the form
\begin{equation}\label{H}
    H = \sum_{\a=1}^L \omega_{\a}b^{\dagger}_{\a}b_{\a}
\end{equation}
with $[b_{\a},b^{\dagger}_{\b}]=\delta_{\a\b} $ and a general
linear mode transformation

\begin{eqnarray}\label{ST}
    a_i  &=&\sum_{\a=1}^L(S_{i\a}^*b_{\a} + T_{i\a}b_{\a}^{\dagger}),\cr
    {a_i}^{\dagger}  &=&\sum_{\a=1}^L(S_{i\a}b_{\a}^{\dagger} +
    T_{i\a}^*b_{\a}).
\end{eqnarray}
To ensure the correct commutation relations between the new modes,
the matrices $S$ and $T$ should satisfy:
\begin{equation}\label{ST2}
    [S^{\dagger},T]=0\ ,\ \ S^{\dagger}S-TT^{\dagger}=I.
\end{equation}
 We ask how much entanglement exists between modes $k$ and $l$.
This requires that we calculate the reduced density matrix of
these two modes, $\rho_{kl}:=tr_{\widehat{k,l}}(\rho_{{\rm {th}}})
$ where $\rho_{{\rm{th}}}=\frac{1}{Z}e^{-\beta H}$,
$Z=tr(e^{-\beta H})$, and $\widehat{k,l}$ means that the modes $k$
and $l$ are excluded when taking the trace.\\
We will show that the reduced density matrix of any two modes, is
always a gaussian state(?). We obtain the condition under which the
transformations (\ref{ST}) can not produce entanglement. We then
consider an example which is a prototype of a wider class and
calculate exactly the entanglement between the two modes and its
dependence on temperature, specifically we obtain the threshold
temperature
above which quantum correlations are destroyed.\\
We begin by collecting the necessary ingredients about gaussian
states that we need in the sequel. In the Hilbert space of two
harmonic oscillators a density matrix $\rho$ is called a two mode
gaussian state if its characteristic function, defined as
$$ C(z_1^*,z_1,z_2^*,z_2)=
tr(e^{z_1a_1^{\dagger}-z_1^*a_1+z_2a_2^{\dagger}-z_2^*a_2}\rho),
$$ is a gaussian function. This can be written
compactly as
\begin{equation}\label{form} C(z_1^*,z_1,z_2^*,z_2)=
e^{-\frac{1}{2}{\bf z}^{\dagger}M{\bf z}},
\end{equation}
where ${\bf z}^{\dagger}=\left(\begin{array}{cccc}
                        z_1^* & z_1 & z_2^* & z_2 \\
                               \end{array} \right)$
and $M$, parameterized as
\begin{equation}\label{M}
    M= \left(
\begin{array}{cc}
  {\mathbf{\a}} & {\mathbf{\gamma}} \\
  {\mathbf {\gamma}}^{\dagger} & {\mathbf {\beta}} \\
\end{array}
    \right)\equiv
    \left(\begin{array}{cccc}
       n_1 & m_1 & m_s & m_c \\
              m_1^* & n_1 & m_c^* & m_s^* \\
              m_s^* & m_c & n_2 & m_2 \\
              m_c^* & m_s & m_2^* & n_2 \\
            \end{array}\right)
\end{equation}
is called the covariance matrix \cite{tut}. \\
This matrix encodes all the correlations in the form
$M_{rs}=\frac{(-1)^{r+s}}{2}\la
 v_r{v^{\dagger}}_s+{v^{\dagger}}_sv_r\ra$,
where $v^{\dagger}=(a_1^{\dagger}, a_1, a_2^{\dagger}, a_2)$ and
$v:=(a_1, a_1^{\dagger},a_2, a_2^{\dagger})^T$.\\ The conditions of
separability of the two modes have been studied in a number of works
\cite{giedke, oliv}. In particular it has been shown that by a
canonical transformation the covariance matrix can be put that when
the covariance for a class of in which $m_c=m_1=m_2=0$. The
conditions of separability then simplify to the following
inequalities \cite{oliv}:
\begin{equation}\label{sep1}
n_1\geq \frac{1}{2} \ \ \ \ {\rm and} \ \ \ \
(n_1-\frac{1}{2})(n_2-\frac{1}{2})\geq |m_s|^2.
\end{equation}
The above conditions only determine the separability of a gaussian
state and not the amount of its entanglement. For a class of
gaussian states invariant under the interchange of the two modes
and called symmetric states, one can actually calculate in closed
form the entanglement of formation \cite{cirac}. The closed
formula is given in terms of the covariance matrix $\Gamma$
defined as $\Gamma_{rs} = \la \eta_r\eta_s+\eta_s\eta_r\ra $,
where $\eta:=(x_1, p_1, x_2, p_2)$ ,
$x_r:=\frac{1}{\sqrt{2}}(a_r+a_r^{\dagger})$ and
$p_r:=\frac{i}{\sqrt{2}}(a_r^{\dagger}-a_r),\  r = 1,2$. For the
covariance matrix $M$ given in (\ref{M}), the covariance matrix
$\Gamma$ will have the form:
\begin{equation}\label{gamma2}
    \Gamma=2\left(
\begin{array}{cc}
  Q & 0 \\
  0 & Q \\
\end{array}\right)M\left(
\begin{array}{cc}
  Q' & 0 \\
  0 & Q' \\
\end{array}\
\right),
\end{equation}
where $Q=\frac{1}{\sqrt{2}}\left(
\begin{array}{cc}
  -1 & 1 \\
  i & i \\
\end{array}
\right)$ and $Q'=\frac{1}{\sqrt{2}}\left(
\begin{array}{cc}
  -1 & -i \\
  1 & -i \\
\end{array}
\right)$. \\
For symmetric gaussian states (those for which $\det {\mathbf
{\alpha}}=\det{\mathbf {\beta}}$) one can apply local symplectic
transformation \cite{giedke2, illum} and without changing their
entanglement put their covariance matrix into the normal form
\begin{equation}\label{normal}
    \Gamma = \left(\begin{array}{cccc}
                        n& 0 & k_x & 0 \\
                        0 & n & 0 & -k_p \\
                        k_x & 0 & n & 0 \\
                        0 & -k_p & 0 & n \\
                      \end{array} \right),
\end{equation}
where $k_x\geq k_p\geq 0 $. These new parameters are derivable
from the symplectic invariants of the matrix $M$:
$$ n^2=4\det{\mathbf {\a}},\ \ k_xk_p=4|\det{\mathbf {\gamma}}|,$$
$$ (n^2-k_x^2)(n^2-k_p^2)=4\det M.$$

The entanglement of formation of these symmetric states is then
given by
\begin{equation}\label{f}
  E_f(\rho):= c_+\log c_+ - c_-\log c_- ,
\end{equation}
where $c_{\pm}=\frac{(1\pm \Delta)^2}{4\Delta}$, and
\be\label{delta}
 \Delta = min\left(1,\delta:=\sqrt{(n-k_x)(n-k_p)}\right). \ee
The state is entangled only when $\Delta<1.$\\
Finally we note that for a state like
$\rho_{\a}:=\frac{1}{Z}e^{-\beta\omega_{\a}
b_{\a}^{\dagger}b_{\a}}$ where $b_{\a}$ and $b_{\a}^{\dagger}$ are
the usual harmonic oscillator operators,
$[b_{\a},b_{\a}^{\dagger}]=1$, and $Z=(1-e^{-\beta
\omega_{\a}})^{-1}$, the single mode characteristic function $
C_{\a}(z^*,z):=tr(e^{zb^{\dagger}-z^*b}\rho_{\a})$ is found to be
\begin{equation}\label{cha}
C_{\a}(z^*,z)=e^{\frac{-|z|^2}{2}\coth \frac{\beta
\omega_{\a}}{2}}.
\end{equation}
One way for doing this calculation is to expand the trace in the
eigenstates of $b^{\dagger}b$. Then by using the properties of
coherent states $|z\ra:= e^{zb^{\dagger}}|0\ra$, (i.e.
$b|z\ra=z|z\ra, \la z|\gamma\ra = e^{z^*\gamma}$ and
$\frac{1}{\pi}\int d\gamma d\gamma^* e^{-|\gamma|^2}|\gamma\ra\la
\gamma| = I $), one writes \ba && \la
n|e^{-z^*b}e^{zb^{\dagger}}|n\ra= \frac{1}{n!}\la
-z|b^n{b^{\dagger}}^n|z\ra\cr &=& \int d\gamma d\gamma^*
|\gamma|^{2n}e^{-z^*\gamma+\gamma^*z-|\gamma|^2}. \ea By summing
over $n$ and performing the resulting gaussian integration one
arrives at the stated result. Also for any two commuting modes
\be\label{c22} C_{\a\b}(z_1^*, z_1, z_2^*,z_2) =
C_{\a}(z_1^*,z_1)C_{\b}(z_2^*,z_2).\ee
 This completes
our short review of gaussian states.\\ We now consider our mode
transformations (\ref{ST}).
 To
express the covariance matrix  of any two modes, say the modes $k$
and $l$ it is useful to introduce a compact notation. Let
\begin{equation}\label{S}
    |S_k\ra:= \left(\begin{array}{c}
            S_{k1} \\
                  S_{k2} \\
                    \cdot \\
                    S_{kL} \\
                  \end{array}\right) \ \ , \ \
                   |T_k\ra:= \left(\begin{array}{c}
            T_{k1} \\
                  T_{k2} \\
                    \cdot \\
                    T_{kL} \\
                  \end{array}\right) ,
\end{equation}
and define the positive inner product between any two such vectors
as
 \begin{equation}  \la X|Y\ra := \sum_{\a=1}^L \coth
\frac{\beta
    \omega_{\a}}{2}X_{\a}^*Y_{\a}.
    \ee

The characteristic function of the two modes is given by
\begin{equation}\label{ckl}
    C_{kl}(z_1^*,z_1, z_2^*, z_2) =
    tr(e^{z_1a_k^{\dagger}-z_1^*a_k+z_2a_l^{\dagger}-z_2^*a_l}\otimes_{\a=1}^L\rho_{\a}).
    \end{equation}
By noting from (\ref{H}) that $\rho_{\rm{th}}=\bigotimes \rho_{\a}
$, and inserting (\ref{ST}) in (\ref{ckl}), rearranging the terms
in the exponential and using (\ref{cha}) and (\ref{c22}) we find
$$    C_{kl}=e^{-\frac{1}{2}\sum_{\a=1}^L|z_1S_{k\a}-z^*_1T_{k\a}+z_2S_{l\a}-z^*_2T_{l\a}
      |^2\cosh \frac{\beta \omega_{\a}}{2}}.
  $$
Comparing with (\ref{form}, \ref{M}) we read the following matrix
elements of the covariance matrix: ($ r=k,l$):
\begin{eqnarray}\label{par}
    n_r &=& \frac{1}{2}( \la S_r|S_r\ra +\la T_r|T_r\ra) \ , \ m_r = -  \la
    S_r|T_r\ra , \cr
    m_s &=& \frac{1}{2}( \la S_k|S_l\ra +\la T_l|T_k\ra), \cr
    m_c &=&
    \frac{-1}{2}( \la S_k|T_l\ra +\la S_l|T_k\ra).
    \end{eqnarray}
We now prove that in any transformation which leaves the vacuum
invariant, the new modes are disentangled. Any such transformation
is one in which $T_{i\a}=0\ \ \forall \a$. To show this we note
that in this case the covariance matrix defined in (\ref{M}) will
have the parameters $m_1 = m_2 = m_c = 0 $
$$ n_1 = \frac{1}{2}\la S_k|S_k\ra \
,\ \ n_2 = \frac{1}{2}\la S_l|S_l\ra \ , \ \ m_s = \frac{1}{2}\la
S_k|S_l\ra . $$ Therefore in view of (\ref{sep1}) the state will
be separable if and only if $$\la S_k|S_k\ra \geq 1 \ , \ (\la
S_k|S_k\ra -1)(\la S_l|S_l\ra -1)\geq |\la S_k|S_l\ra|^2. $$ Due
to (\ref{ST2}) we know that $\sum_{\a=1}^L|S_{k\a}|^2=1 $ and
$\sum_{\a=1}^LS_{k\a}^*S_{l\a}=0$. Since $\coth
\frac{\beta\omega}{2}\geq 1$, it is obvious that the first
inequality is satisfied. If we know introduce a new inner product
as $$(X|Y):= \sum_{\a=1}^LX^*_{\a}Y_{\a}(\coth \frac{\beta
\omega_{\a}}{2}-1),$$ the second condition takes the form \be
(S_k|S_k)(S_l|S_l)\geq |(S_k|S_l)|^2, \ee
 which is satisfied by the Cuachy Schawrz inequality. This completes the proof.\\
The above argument is also true for the transformations for which
$S=0$ and $T\ne 0$. These new modes can also be called vacuum
preserving by renaming the new creation and annihilation
operators. The only transformations which can produce entanglement
are those for which neither $S$ nor $T$ vanish. Thus for systems
of identical particles, these kinds of transformations play the
role of non-local transformation which can produce entanglement.
\\

\section{An Example}
We will now consider such a transformation and for this purpose we
choose an example which shows that our formalism is also applicable
to systems of identical but localized and distinguishable bosonic
particles. Consider two particles (atoms) oscillating around their
equilibrium positions modeled by a mass-spring system with a
Hamiltonian
$$    H =
    \frac{p_1^2}{2}+\frac{p_2^2}{2}+\frac{1}{2}x_1^2+\frac{1}{2}x_2^2+\frac{1}{2}\omega_0^2(x_1-x_2)^2,
$$ where $x_i$ and $p_i$ denote the canonical
coordinate and momentum of the $i$-th particle. Note that the
particles are localized and distinguishable in this case. We want
to calculate the thermal entanglement of these two atoms and study
its dependence on temperature. In particular we want to see if
there is a threshold temperature above which entanglement
vanishes. This type of study has been intensively carried out for
spins systems \cite{niel,arn, ocw, zan2} and to our knowledge this
is the first time where thermal entanglement of continuous degrees
of freedom is being studied. \\
We first find the the normal coordinates of the system; $$ X_{1} =
\frac{1}{\sqrt{2}}(x_1+x_2),\ \ \ X_{2}=:
\frac{1}{\sqrt{2}}(x_1-x_2), $$
  $$ P_{1} := \frac{1}{\sqrt{2}}(p_1+p_2),\ \ \ P_{2}=:
    \frac{1}{\sqrt{2}}(p_1-p_2),$$
and the oscillator modes $ b_{1}
:=\frac{1}{\sqrt{2}}(X_{1}+iP_{1})\ \ $ and $
    b_{2} :=\frac{1}{\sqrt{2\omega}}(\omega X_{2}+iP_{2}),$ where
    $\omega:=\sqrt{1+2\omega_0^2}$.
These modes diagonalize the Hamiltonian
$$    H = b_1^{\dagger}b_{1} + \omega b_2^{\dagger}b_{2},
$$
where we have ignored an overall constant. \\
For calculating the thermal density matrix of the two particles,
we proceed as before to determine the characteristic functions of
the two modes where mode now means the degree of freedom of each
individual atom. That is we have $
    a_1 = \frac{1}{\sqrt{2}}(x_1+ip_1)\ \ $ and $\ \ a_2 =
    \frac{1}{\sqrt{2}}(x_2+ip_2)$.
The relation of the new modes with the old ones turns out to be
$$
    a_1 = \frac{1}{\sqrt{2}}\left(b_1 + \xi_{+}b_2+\xi_{-}b^{\dagger}_2\right), $$
$$  a_2 = \frac{1}{\sqrt{2}}\left(b_1 - \xi_{+}b_2-\xi_-
b^{\dagger}_2\right),
$$ where $\xi_{\pm} =
\frac{1}{2}\left(\sqrt{\omega^{-1}} \pm \sqrt{\omega}\right)$.
 From this equation and (\ref{ST},\ref{S}) we find
that
\begin{equation}\label{sss1}
    |S_1\ra = \frac{1}{\sqrt{2}}\left(\begin{array}{c}
                                        1 \\
                                        \xi_+ \\ \end{array}\right)\ \ \ ,\ \ \
|T_1\ra = \frac{1}{\sqrt{2}}\left(\begin{array}{c}
                                        0 \\
                                       \xi_- \\ \end{array}\right),
\end{equation}

and
\begin{equation}\label{sss2}
    |S_2\ra = \frac{1}{\sqrt{2}}\left(\begin{array}{c}
                                        1 \\
                                        -\xi_+ \\ \end{array}\right)\ \ \ , \ \ \
|T_2\ra = \frac{1}{\sqrt{2}}\left(\begin{array}{c}
                                        0 \\
                                        -\xi_- \\ \end{array}\right).
\end{equation}

In order to simplify the notation lets us set  $u:=\coth
\frac{\beta}{2}$ and $v:=\coth\frac{\beta \omega}{2}$,
$a:=\frac{1}{2}(\omega^{-1}+\omega)v$, and
$b:=\frac{1}{2}(\omega^{-1}-\omega)v$.
 With these conventions we
will find from (\ref{sss1},\ref{sss2}) and (\ref{par}) that:
$$M= \frac{1}{4}\left(
\begin{array}{cccc}
  u+a & -b & u-a & b \\
  -b & u+a & b & u-a \\
  u-a & b & u+a & -b \\
  b & u-a & -b & u+a \\
\end{array}
\right). $$
  Using the relations (\ref{gamma2})
 we find the following form for the matrix $\Gamma $ :

$$ \Gamma:= \frac{1}{2}\left(
\begin{array}{cccc}
  u+a+b & 0 & u-a-b &0 \\
  0 & u+a-b & 0 & u-a+b \\
  u-a-b & 0 & u+a+b & 0 \\
  0 & u-a+b & 0 & u+a-b \\
\end{array}
\right). $$ This is not yet the final symmetric form of the matrix
$\Gamma$ as in (\ref{normal}),
 from which we can calculate the entanglement. For
this last step we need the parameters $n$, $k_x$ and $k_p$ which
can be derived from the symplectic invariants of $\Gamma $.
Actually it is simpler to do a canonical transformation
$x_i\rightarrow \alpha x_i, \ \ p_i\rightarrow \frac{1}{\alpha}
p_i $ with $\a := (\frac{u+a-b}{u+a+b})^{\frac{1}{4}}$ to put this
matrix in the symmetric form (\ref{normal}) and read the
parameters $ n, k_x $ and $k_p$. The result is: \ba n&=&
\frac{1}{2}\sqrt{(u+a)^2-b^2}, \cr k_x&=&\frac{1}{2}(u-a-b)
(\frac{u+a-b}{u+a+b})^{\frac{1}{2}},\cr -k_p&=&\frac{1}{2}(u-a+b)
(\frac{u+a+b}{u+a-b})^{\frac{1}{2}}, \ea leading to
$$  n-k_x=(a+b)\sqrt{\frac{u+a-b}{u+a+b}},\ \ \ n-k_p=
    u\sqrt{\frac{u+a+b}{u+a-b}}.
$$ From this last result and the definition of
$\Delta$ in
 (\ref{delta})
we find the condition of entanglement of the two atoms
\begin{equation}\label{atoms}
    \delta^2= u(a+b)\equiv \frac{\coth \frac{\beta}{2} \coth \frac{\beta \omega}{2}}{\omega} <
    1.
\end{equation}
Since $\omega > 1$, this inequality can be satisfied below a
threshold temperature $T_c$ obtained by setting
$\delta^2(T_c,\omega) = 1 $. Inserting the value of $\delta$ from
(\ref{atoms}) in (\ref{delta}) and then using (\ref{f}) we obtain
the entanglement between the two atoms as a function of
temperature and frequency. The result is plotted in figure 1. The
entanglement at zero temperature is obtained by inserting the
value of delta in this limit, $\Delta =\frac{1}{\sqrt{\omega}}$,
in (\ref{f}). This leads to $$
    E_{max}\equiv E(T=0) = x\ln x - (x-1)\ln (x-1),
$$
where $x:= \frac{(1+\sqrt{\omega})^2}{4\sqrt{\omega}}$. This is an
increasing function of $\omega$ as shown in figure 1.

\begin{figure}[t]
\centering
\includegraphics[width=8cm,height=8cm,angle=0]{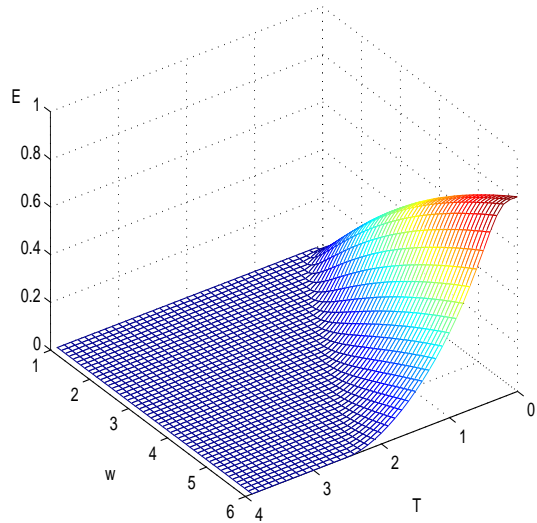}
\caption{Entanglement of the two oscillators as a function of the
natural frequency and temperature. The threshold temperature
increases almost linearly with frequency.}
\end{figure}
The maximum entanglement behaves like $\ \ (\frac{1}{16\ln
2}+\frac{1}{4}-\frac{1}{8}\log_2(w-1))(w-1)^2\ $ for small
frequencies $\ \omega \approx 1 (\omega_0<< 1)\ $ and like $\ \
\frac{1}{\ln 2}-2+\frac{1}{2}\log_2\omega\ $ for
large frequencies $\ \omega>> 1$. \\

In summary we have set up an easy formalism for calculating the
thermal entanglement of arbitrary bosonic modes in a rather large
class of problems. For example the formalism can be applied to
arbitrary lattices of bosonic modes  to see how thermal fluctuations
affect frustration of entanglement \cite{cirac3}, or to a chain of
coupled oscillators. In the later case the entanglement can be
obtained as a function of both the temperature and the distance
between the particles. This is in contrast to the spin systems where
due to the complicated nature of their spectrum this later
dependence can not be obtained except only at low
temperatures and under certain assumptions \cite{ak}. \\
I would like to thank V. Karimipour, A. Bayat, I. Marvian, L.
Memarzadeh and A. Sheikhan for very valuable discussions.

{}

\end{document}